\documentclass{ws-ijmpa}
\usepackage[super,compress]{cite}
\usepackage{graphicx}
\begin{document}
\markboth{Z.-H. Weng}{Contrastive analysis of two energy gradients}

\catchline{}{}{}{}{}
%
%%%%%%%%%%%%%%%%%%%%%%%%%%%%%%%%%%%%%%%%%%%%%%%%%%%%%%%%%%%%%%%%%%%%

\title{Contrastive analysis of two energy gradients in the ultra-strong magnetic fields
%\footnote{
%For the title, try not to use more than
%3 lines. Typeset the title in 10 pt roman, uppercase and
%boldface.}
}

\author{Zi-Hua Weng
%\footnote{
%Typeset names in 8 pt roman, uppercase. Use the footnote to indicate the
%present or permanent address of the author.}
}

\address{School of Aerospace Engineering, Xiamen University, Xiamen, China
\\
College of Physical Science and Technology, Xiamen University, Xiamen, China
%\footnote{
%State completely without abbreviations, the affiliation and
%mailing address, including country. Typeset in 8 pt italic.}
\\
xmuwzh@xmu.edu.cn}

%\author{Second Author}
%
%\address{Group, Laboratory, Address\\
%City, State ZIP/Zone, Country\\
%second\_author@domain\_name}

\maketitle

\begin{history}
\received{Day Month Year}
\revised{Day Month Year}
\end{history}

\begin{abstract}
  The paper aims to apply the complex-octonions to explore the variable gravitational mass and energy gradient of several particles in the external ultra-strong magnetic fields. J. C. Maxwell was the first to introduce the algebra of quaternions to study the physical properties of electromagnetic fields. Some scholars follow up this method in the field theories. Nowadays, they employ the complex-octonions to analyze simultaneously the physical quantities of electromagnetic and gravitational fields, including the field potential, field strength, field source, linear momentum, angular momentum, torque, and force. When the octonion force is equal to zero, it is able to deduce eight independent equilibrium equations, especially the force equilibrium equation, precession equilibrium equation, mass continuity equation, and current continuity equation. In the force equilibrium equation, the gravitational mass is variable. The gravitational mass is the sum of the inertial mass and a few tiny terms. These tiny terms will be varied with not only the fluctuation of field strength and of potential energy, but also the spatial dimension of velocity. The study reveals that it is comparatively untoward to attempt to measure directly the variation of these tiny terms of gravitational mass in the ultra-strong magnetic field. However it is not such difficult to measure the energy gradient relevant to the variation of these tiny terms of gravitational mass. In the complex-octonion space, the gravitational mass is a sort of variable physical quantity, rather than an intrinsic property of any physical object. And this inference is accordant with the academic thought of `the mass is not an intrinsic property any more' in the unified electro-weak theory.

\keywords{energy gradient; gravitational mass; ultra-strong magnetic field; E\"{o}tv\"{o}s experiment; octonion.}

\end{abstract}

\ccode{PACS numbers: 04.80.Cc, 04.50.-h, 11.10.Kk, 02.10.De}

%\tableofcontents

\section{Introduction}

Why do we need two different kinds of masses, gravitational mass and inertial mass? In the force equilibrium equation, there are the gravitational force and inertial force and so forth. The gravitational mass relates with the gravitational force, while the inertial mass is relevant to the inertial force. Why must the gravitational mass be equal to the inertial mass, when the force equilibrium equation includes the contributions from various types of force terms? After the mass has been transformed from an intrinsic property into one dynamic property in the unified electro-weak theory, could the magnitude of gravitational mass not be equal to that of inertial mass? The gravitational mass and inertial mass may be two distinct physical quantities, although two of them seem to be quite similar on the face of it. Since a long time ago, these puzzles have been intriguing and bewildering the scholars. Until recently, the emergence of electromagnetic and gravitational theories, described with the complex-octonions, replies to a part of these conundrums. Under some extreme conditions, such as the ultra-strong magnetic fields, the gravitational mass is variable. The magnitude of gravitational mass may be deviated from that of inertial mass to a certain extent. It will be beneficial to further understand the physical property of gravitational mass.

In 1736, L. Euler developed and clarified the Newton's second law of motion. On the basis of the mathematical expression, the ratio of force to acceleration is defined as the inertial mass. In 1879, W. Thomson and P. G. Tait introduced the gravitational mass. Based on the universal gravitation, it is able to define clearly the gravitational mass. By all appearances, the physical properties of these two types of masses are distinct, and the measurement methods of two masses are disparate either. The scholars have been suspecting the equivalence of two masses for a long time. Since the beginning of the twentieth century, the controversy about the equivalence of two masses has never stopped, although there are a part of experimental results. Recently, a new point of view becomes known suddenly. That is, the Higgs mechanism \cite{aranda} in the unified electro-weak theory reveals that the mass is the dynamic property rather than an intrinsic property.

In the Newtonian theory, the scholars explore the equivalence principle, E\"{o}tv\"{o}s experiment, gravitational mass, and energy gradient and so forth. Damour \cite{GA2} discussed the Equivalence Principle violation phenomenology of dilaton-like models. Schlamminger \emph{et al.} \cite{GA15} applied a continuously rotating torsion balance instrument, measuring the acceleration difference of beryllium and titanium test bodies. Moffat \cite{GA16} proposed to apply the E\"{o}tv\"{o}s experiment to test the weak equivalence principle in the satellite. Wei \emph{et al.} \cite{E15} found a way of testing the Equivalence Principle, by means of the observed time delays between different energy bands from the blazars. Mohapi \emph{et al.} \cite{E16} proposed a general tensor-scalar theory, testing the equivalence principle in the dark sector. Han \emph{et al.} \cite{E17} considered an experiment scheme for a space free-fall based test of the equivalence principle. Reasenberg \cite{E18} applied a new class of test masses to a Galilean test of the equivalence principle. Haghi \emph{et al.} \cite{E19} found that the equivalence principle will be violated, in case the gravitational dynamics is impacted by the external gravitational field. Overduin \emph{et al.} \cite{E21} revealed the equivalence principle will be violated, in terms of the observational uncertainties in the positions and motions of solar-system bodies. Hardy \emph{et al.} \cite{E23} focused on testing the equivalence principle, by means of the precise measurement on the microsatellite. Freire \emph{et al.} \cite{E26} presented an equivalence principle violation test, on the basis of measuring the variation of the orbital eccentricity. Barrett \emph{et al.} \cite{E27} proposed two methods to extract the differential phase between dual-species atom interferometers, testing the weak equivalence principle. Donoghue \emph{et al.} \cite{E29} demonstrated the long-distance corrections at one loop, leading to the quantum violations of some classical consequences of the equivalence principle. Zhou \emph{et al.} \cite{E30} reported on a test of the weak equivalence principle, by means of a simultaneous dual-species atom interferometer. Tarallo \emph{et al.} \cite{E31} notified a test of the equivalence principle, measuring the acceleration of two isotopes of strontium atoms in the Earth's gravity field. Williams \emph{et al.} \cite{E36} studied the quantum test of the equivalence principle on the International Space Station. Bjerrum-Bohr \emph{et al.} \cite{E39} applied the modern methods to the quantum gravity at low energy, challenging the equivalence principle. Gnatenko \cite{E42} notified a non-uniform gravitational field to violate the weak equivalence principle.

Making a detailed comparison and analysis of the preceding studies, a few primal problems, associated with the `gravitational mass and energy gradient' in the Newtonian theory, are found as follows.

1) Vulnerable postulate. In the Newtonian theory, the gravitational mass is supposed to be invariable, and is irrelevant to the fluctuation of the energy either. However, the suspicion about the inference, derived from the experimental results of the E\"{o}tv\"{o}s experiment, keeps coming until now. Up to now most of E\"{o}tv\"{o}s experiments are only validated experimentally in the quite weak gravitational field on the earth's surface. Especially the E\"{o}tv\"{o}s experiment has never been verified in the comparatively strong magnetic fields or gravitational fields. The equivalence principle has been becoming a vulnerable point to be criticized. As a result, it is necessary to survey the E\"{o}tv\"{o}s experiment in the ultra-strong magnetic fields.

2) Perplexed astrophysical jets. As one constituent part of the force, the energy gradient in the Newtonian theory is unable to explain the physical phenomena of astrophysical jets. In the galactic and extragalactic systems, the scholars observe the jet phenomena in a large number of celestial bodies. Many observations state that the astrophysical jets are ubiquitous. However, the production mechanism of the astrophysical jets is still an unsolved puzzle in the radio astronomy up to now. The astrophysical jets are challenging the Newtonian theory and other existing theories. The scholars propose a few theoretical models, but these models are unable to explain effectively the dynamic properties of astrophysical jets.

3) Insufficient investigation. Before the emergence of the General Theory of Relativity (GR for short), the Newtonian theory was deemed to be applicable to all cases. This was obviously due to the lack of the in-depth research of the Newtonian theory then. Similarly, the study of the GR is also extremely limited nowadays, so there are some scholars to reckon that the application scope of GR may be unbounded. That is, GR is expected to be applicable to all cases. However, GR is incapable of explaining a few physical phenomena, including the dark matter and astrophysical jets. Consequently, scholars are suspicious of the GR from time to time. They attempt to validate the equivalence principle from various environments, especially the ultra-strong magnetic fields.

Presenting a striking contrast to the above is that it is able to account for several puzzles, which are derived from the `gravitational mass and energy gradient' of the Newtonian theory, attempting to improve the academic thought of the `gravitational mass and energy gradient' to a certain extent, in the gravitational and electromagnetic theories described with the complex-octonions.

J. C. Maxwell was the first to introduce the algebra of quaternions to explore the physical property of electromagnetic fields. His method edifies the other scholars to apply the quaternions and octonions to study the gravitational fields, electromagnetic fields, GR, and astrophysical jets and so forth. When a part of coordinate values are complex numbers, the quaternions and octonions are called as the complex-quaternions and complex-octonions respectively \cite{weng1}.

Some scholars make use of the quaternions and octonions to describe the electromagnetic field equations, gravitational field equations, GR, and quantum mechanics and so forth. Castro \cite{OB4} presented a nonassociative octonionic gauge field theories, on the basis of a ternary bracket. Ludkowski \emph{et al.} \cite{OB5} applied the super-differential operators over the quaternion skew field and octonion algebra to study the spectral theory. Pushpa \emph{et al.} \cite{OB6} made use of the relation between quaternion basis elements with Pauli matrices and octonions, discussing the grand unified theories. Chanyal \emph{et al.} \cite{OB7} applied the octonion variables to reformulate the generalized field equation of dyons. Tsagas \cite{OB13} studied the electromagnetic fields in the curved spacetimes, deducing the wave equations for the associated electric and magnetic components. Morita \cite{O21} considered the relations among the complex quaternions, restricted Lorentz group, and Dirac theory. Rawat \emph{et al.} \cite{O26} shown that the quaternionic formulation remains invariant under the quaternion transformations, where real and imaginary parts represent gravitation and electromagnetism respectively. Mironov \emph{et al.} \cite{O27} derived the relations for energy, momentum, and Lorentz invariants of the electromagnetic field, by means of the algebra of octonions. Demir \emph{et al.} \cite{O28} formulated the generalized field equations of linear gravity, on the basis of the algebra of octonions. Gogberashvili \cite{O29} applied the octonions to explore the Dirac's operator and Maxwell's equations. Negi \emph{et al.} \cite{O30} made an attempt to reformulate the generalized field equation of dyons in terms of octonion variables. Furui \cite{O31} applied the triality symmetry of octonions to discuss the axial current and axial anomaly.

In the paper, applying of the complex octonions is able to describe the physical quantities in the electromagnetic and gravitational fields simultaneously, including the field potential, field strength, field source, linear momentum, angular momentum, torque, and force and so forth. The electromagnetic and gravitational theories, described with the complex octonions, can resolve several problems remaining to be solved in the Newtonian theory.

1) Alterable gravitational mass. In the field theory described with the octonions (octonion field theory for short, temporarily), the gravitational mass is variable, and is relevant to the fluctuation of the energy. The comparatively strong electromagnetic strength or/and gravitational strength may induce the variation of gravitational mass. In the E\"{o}tv\"{o}s experiment, in case the distribution of energy is non-uniform, the energy gradient will impact the state of force equilibrium. Moreover, the comparatively strong field strength is capable of amplifying the influence of energy gradient, violating severely the E\"{o}tv\"{o}s experiment in the non-uniform and strong field strength.

2) Dynamic of astrophysical jets. As one component of the force, the energy gradient associates with the variable gravitational mass. The strength gradient is dependent of neither the direction of field strength nor the mass and electric charge for the test particle. In the octonion field theory, when the strength gradient is considered as the thrust of astrophysical jets, one can infer several movement features of astrophysical jets \cite{weng2}. Especially, the strength gradient is able to unpuzzle the mechanical features of astrophysical jets, including the bipolarity, matter ingredient, precession, collimation, stability, continuing acceleration and so forth.

3) Limited scope of application. In the octonion field theory, the physical property of gravitational mass is distinct from that of inertial mass. In the strict sense, only if the gravitational strength and electromagnetic strength are all zero, the gravitational mass may be equal to the inertial mass approximately. The study of gravitational mass and strength gradient reveal that the application scope of GR is restricted, that is, GR is unsuitable to all cases. And GR is just fit for the exceptional circumstances that the variation of gravitational mass is zero.

In the paper, making use of the quaternion operator and octonion field potential (in Section 2), it is able to define the octonion field strength, field source, linear momentum, angular momentum, torque, and force and so forth. When the octonion force is equal to zero, one can deduce the force equilibrium equation, precession equilibrium equation, mass continuity equation, and current continuity equation and so on. The force equilibrium equation reveals that there is a tiny magnitude of deviation between the gravitational mass and inertial mass. It should be noted that this tiny deviation will result in the significant fluctuation of energy gradient, impacting directly the E\"{o}tv\"{o}s experiment and relevant inferences. In the ultra-strong magnetic fields, the paper launches the contrastive analysis of the energy gradients between the Newtonian theory and octonion field theory. It is predicted that the gravitational mass may possess a few new physical properties, deepening further the understanding of the gravitational mass.

\section{Force and energy}

According to the academic thought of R. Descartes and M. Faraday, the space is only the extension of the fundamental field (such as, electromagnetic or gravitational field) and does not claim existence on its own. Further each fundamental field extends its individual space. These spaces are quite similar but independent to each other. According to three postulates of the octonion field theory (see Ref.[20]), the space extended from either the electromagnetic field or gravitational field is the quaternion space. Moreover, the quaternion space $\mathbb{H}_e$ for the electromagnetic field is independent to the quaternion space $\mathbb{H}_g$ for the gravitational field. The two quaternion spaces, $\mathbb{H}_e$ and $\mathbb{H}_g$ , are perpendicular to each other, and they can be combined together to become an octonion space. In the octonion space $\mathbb{O}$ , it is able to describe simultaneously the physical properties of electromagnetic and gravitational fields.

Someone may think all of `one-dimensional coordinate axes' are the same. But it's not a fact. On the one hand, the coordinate ratios of some one-dimensional coordinate axes may be different from each other. On the other hand, a few coordinate axes may be perpendicular to each other. These deductions of one-dimensional coordinate axes can be extended into the multi-dimensional spaces, because the one-dimensional coordinate axis is considered as a one-dimensional space. In other words, several quaternion spaces should be not the same. Not only may the coordinate ratio of quaternion spaces be different from each other, but also the quaternion spaces can be perpendicular to each other.

\subsection{Angular momentum}

In the quaternion space $\mathbb{H}_g$ for the gravitational field, the basis vector is $\emph{\textbf{i}}_j$ , the radius vector is $\mathbb{R}_g = i r_0 \emph{\textbf{i}}_0 + \Sigma r_k \emph{\textbf{i}}_k$ , the velocity is $\mathbb{V}_g = i v_0 \emph{\textbf{i}}_0 + \Sigma v_k \emph{\textbf{i}}_k$ . The gravitational potential is $\mathbb{A}_g = i a_0 \emph{\textbf{i}}_0 + \Sigma a_k \emph{\textbf{i}}_k$ , the gravitational strength is $\mathbb{F}_g = f_0 \emph{\textbf{i}}_0 + \Sigma f_k \emph{\textbf{i}}_k$, the gravitational source is $\mathbb{S}_g = i s_0 \emph{\textbf{i}}_0 + \Sigma s_k \emph{\textbf{i}}_k$ . Meanwhile, in the 2-quaternion space $\mathbb{H}_e$ for the electromagnetic field, the basis vector is $\emph{\textbf{I}}_j$ , the radius vector is $\mathbb{R}_e = i R_0 \emph{\textbf{I}}_0 + \Sigma R_k \emph{\textbf{I}}_k$ , the velocity is $\mathbb{V}_e = i V_0 \emph{\textbf{I}}_0 + \Sigma V_k \emph{\textbf{I}}_k$ . The electromagnetic potential is $\mathbb{A}_e = i A_0 \emph{\textbf{I}}_0 + \Sigma A_k \emph{\textbf{I}}_k$ , the electromagnetic strength is $\mathbb{F}_e = F_0 \emph{\textbf{I}}_0 + \Sigma F_k \emph{\textbf{I}}_k$, the electromagnetic source is $\mathbb{S}_e = i S_0 \emph{\textbf{I}}_0 + \Sigma S_k \emph{\textbf{I}}_k$ . Herein $\emph{\textbf{r}} = \Sigma r_k \emph{\textbf{i}}_k$ . $\emph{\textbf{v}} = \Sigma v_k \emph{\textbf{i}}_k$ . $\emph{\textbf{a}} = \Sigma a_k \emph{\textbf{i}}_k$. $\emph{\textbf{f}} = \Sigma f_k \emph{\textbf{i}}_k$ . $\emph{\textbf{s}} = \Sigma s_k \emph{\textbf{i}}_k$ . The quaternion operator is $\lozenge = i \emph{\textbf{i}}_0 \partial_0 + \Sigma \emph{\textbf{i}}_k \partial_k$ . $\nabla = \Sigma \emph{\textbf{i}}_k \partial_k$ , $\partial_j = \partial / \partial r_j$ . $r_0 = v_0 t$ . $v_0$ is the speed of light, and $t$ is the time. $\textbf{R}_0 = R_0 \emph{\textbf{I}}_0$, $\textbf{V}_0 = V_0 \emph{\textbf{I}}_0$ , $\textbf{A}_0 = A_0 \emph{\textbf{I}}_0$ , $\textbf{F}_0 = F_0 \emph{\textbf{I}}_0$, $\textbf{S}_0 = S_0 \emph{\textbf{I}}_0$ . $\textbf{R} = \Sigma R_k \emph{\textbf{I}}_k$ , $\textbf{V} = \Sigma V_k \emph{\textbf{I}}_k$ , $\textbf{A} = \Sigma A_k \emph{\textbf{I}}_k$ , $\textbf{F} = \Sigma F_k \emph{\textbf{I}}_k$ , $\textbf{S} = \Sigma S_k \emph{\textbf{I}}_k$. $\emph{\textbf{i}}_0 = 1$ , $\emph{\textbf{i}}_k^{~2} = -1$. $\emph{\textbf{I}}_k = \emph{\textbf{i}}_k \circ \emph{\textbf{I}}_0$ . $\emph{\textbf{I}}_j^{~2} = -1$. The symbol $\circ$ denotes the octonion multiplication. $r_j$ , $v_j$ , $a_j$, $s_j$ , $f_0$ , $R_j$, $V_j$, $A_j$, $S_j$, and $F_0$ are all real. $f_k$ and $F_k$ both are complex numbers. $i$ is the imaginary unit. $j = 0, 1, 2, 3$. $k = 1, 2, 3$.

Two orthogonal complex-quaternion spaces, $\mathbb{H}_e$ and $\mathbb{H}_g$ , can be combined together to become one complex-octonion space $\mathbb{O}$ . In the complex-octonion space $\mathbb{O}$, one can define the octonion source $\mathbb{S}$ from the octonion strength $\mathbb{F}$ (Table 1). And it is capable of defining the octonion linear momentum, $\mathbb{P} = \mathbb{P}_g + k_{eg} \mathbb{P}_e$ , from the octonion source $\mathbb{S}$ . Next, the octonion angular momentum, $\mathbb{L}$ , is defined from the octonion linear momentum $\mathbb{P}$ , octonion radius vector, $\mathbb{R} = \mathbb{R}_g + k_{eg} \mathbb{R}_e$  , and octonion integrating function of field potential $\mathbb{X}$ (see Ref.[20] ). Further, the octonion angular momentum $\mathbb{L}$ can be separated into,
\begin{eqnarray}
\mathbb{L} = \mathbb{L}_g + k_{eg} \mathbb{L}_e  ~,
\end{eqnarray}
where the component $\mathbb{L}_g$ is relevant to the angular momentum and so forth, while the component $\mathbb{L}_e$ is related to the electric/magnetic dipole moment and so on. $\mu$ , $\mu_g$ , $\mu_e$ , and $k_{eg}$ are coefficients. $\mathbb{P}_g = \mathbb{S}_g - i \mathbb{F}^\ast \circ \mathbb{F} / ( v_0 \mu_g )$ ; $\mathbb{P}_e = \mu_e \mathbb{S}_e / \mu_g$. $\mathbb{P}_g = i p_0 + \textbf{p}$ ; $\textbf{p} = \Sigma p_k \emph{\textbf{i}}_k$ . $\mathbb{P}_e = i \textbf{P}_0 + \textbf{P}$ ; $\textbf{P} = \Sigma P_k \emph{\textbf{I}}_k$ . $\textbf{P}_0 = P_0 \emph{\textbf{I}}_0$ . $\mathbb{L}_g = L_{10} + i \textbf{L}_1^i + \textbf{L}_1$ . $\textbf{L}_1 = \Sigma L_{1k} \emph{\textbf{i}}_k$, $\textbf{L}_1^i = \Sigma L_{1k}^i \emph{\textbf{i}}_k$ . $\mathbb{L}_e = \textbf{L}_{20} + i \textbf{L}_2^i + \textbf{L}_2$ . $\textbf{L}_2 = \Sigma L_{2k} \emph{\textbf{I}}_k$, $\textbf{L}_2^i = \Sigma L_{2k}^i \emph{\textbf{I}}_k$ . $\textbf{L}_{20} = L_{20} \emph{\textbf{I}}_0$ .
For one single particle, a comparison with the classical field theory states that, $\mathbb{S}_g = m \mathbb{V}_g$ , $\mathbb{S}_e = q \mathbb{V}_e$ . $m$ is the mass density, while $q$ is the density of electric charge. The gravitational constant is $\mu_g < 0$, and the electromagnetic constant is $\mu_e > 0$. $\ast$ indicates the conjugation of octonion. $\star$ is the complex conjugate. $p_j$ , $P_j$ , $L_{1j}$ , $L_{1k}^i$ , $L_{2j}$ , and $L_{2k}^i$ are all real.

\subsection{Force}

In the complex-octonion space $\mathbb{O}$ , the octonion torque $\mathbb{W}$ is defined from the octonion angular momentum $\mathbb{L}$ . And the octonion torque $\mathbb{W}$ is able to be broken into,
\begin{eqnarray}
\mathbb{W} = \mathbb{W}_g + k_{eg} \mathbb{W}_e  ~,
\end{eqnarray}
where the component $\mathbb{W}_g$ associates with the partial derivative of the angular momentum $\mathbb{L}_g$ and so on, while the component $\mathbb{W}_e$ deals with the partial derivative of the electric/magnetic dipole moment $\mathbb{L}_e$ and so forth. $\mathbb{W}_g = i W_{10}^i + W_{10} + i \textbf{W}_1^i + \textbf{W}_1$. $\mathbb{W}_e = i \textbf{W}_{20}^i + \textbf{W}_{20} + i \textbf{W}_2^i + \textbf{W}_2$ . $W_{10}^i$ is the energy. - $\textbf{W}_1^i$ is the torque. $\textbf{W}_1$ is the curl of angular momentum, and $W_{10}$ is the divergence of angular momentum. $\textbf{W}_1 = \Sigma W_{1k} \emph{\textbf{i}}_k$ , $\textbf{W}_1^i = \Sigma W_{1k}^i \emph{\textbf{i}}_k$ . $\textbf{W}_{20} = W_{20} \emph{\textbf{I}}_0$ , $\textbf{W}_{20}^i = W_{20}^i \emph{\textbf{I}}_0$ . $\textbf{W}_2 = \Sigma W_{2k} \emph{\textbf{I}}_k$ , $\textbf{W}_2^i = \Sigma W_{2k}^i \emph{\textbf{I}}_k$. $W_{1j}$ , $W_{1j}^i$ , $W_{2j}$ , and $W_{2j}^i$ are all real.

Similarly, the octonion force $\mathbb{N}$ is defined from the octonion torque $\mathbb{W}$ . And the octonion force $\mathbb{N}$ may be divided into,
\begin{eqnarray}
\mathbb{N} = \mathbb{N}_g + k_{eg} \mathbb{N}_e  ~,
\end{eqnarray}
where
$\mathbb{N}_g = i N_{10}^i + N_{10} + i \textbf{N}_1^i + \textbf{N}_1$ . $\mathbb{N}_e = i \textbf{N}_{20}^i + \textbf{N}_{20} + i \textbf{N}_2^i + \textbf{N}_2$ . $N_{10}$ is the power, including some terms in the mass continuity equation. $\textbf{N}_1$ is the derivative of the torque, and $\textbf{N}_1^i$ is the force. $\textbf{N}_{20}$ covers several terms in the current continuity equation. $\textbf{N}_1 = \Sigma N_{1k} \emph{\textbf{i}}_k$ , $\textbf{N}_1^i = \Sigma N_{1k}^i \emph{\textbf{i}}_k$ . $\textbf{N}_{20} = N_{20} \emph{\textbf{I}}_0$ , $\textbf{N}_{20}^i = N_{20}^i \emph{\textbf{I}}_0$ . $\textbf{N}_2 = \Sigma N_{2k} \emph{\textbf{I}}_k$ , $\textbf{N}_2^i = \Sigma N_{2k}^i \emph{\textbf{I}}_k$ . $N_{1j}$ , $N_{1j}^i$ , $N_{2j}$ , and $N_{2j}^i$ are all real.

\begin{table}[h]
\tbl{Some physical quantities described with the octonions, in the electromagnetic and gravitational fields.}
%\center
{\begin{tabular}{@{}lll@{}}
\hline\hline
physics~quantity      &   definition                                                                        &   decomposition                                        \\
\hline
field~potential       &  $\mathbb{A} = i \lozenge^\star \circ \mathbb{X}  $                                 &   $\mathbb{A} = \mathbb{A}_g + k_{eg} \mathbb{A}_e$    \\
field~strength        &  $\mathbb{F} = \lozenge \circ \mathbb{A}  $                                         &   $\mathbb{F} = \mathbb{F}_g + k_{eg} \mathbb{F}_e$    \\
field~source          &  $\mu \mathbb{S} = - ( i \mathbb{F} / v_0 + \lozenge )^\ast \circ \mathbb{F} $      &   $\mu \mathbb{S} = \mu_g \mathbb{S}_g + k_{eg} \mu_e \mathbb{S}_e - i \mathbb{F}^\ast \circ \mathbb{F} / v_0 $   \\
linear~momentum       &  $\mathbb{P} = \mu \mathbb{S} / \mu_g $                                             &   $\mathbb{P} = \mathbb{P}_g + k_{eg} \mathbb{P}_e$    \\
angular~momentum      &  $\mathbb{L} = ( \mathbb{R} + k_{rx} \mathbb{X} )^\star \circ \mathbb{P} $          &   $\mathbb{L} = \mathbb{L}_g + k_{eg} \mathbb{L}_e$    \\
octonion~torque       &  $\mathbb{W} = - v_0 ( i \mathbb{F} / v_0 + \lozenge ) \circ \mathbb{L} $           &   $\mathbb{W} = \mathbb{W}_g + k_{eg} \mathbb{W}_e$    \\
octonion~force        &  $\mathbb{N} = - ( i \mathbb{F} / v_0 + \lozenge ) \circ \mathbb{W} $               &   $\mathbb{N} = \mathbb{N}_g + k_{eg} \mathbb{N}_e$    \\
\hline\hline
\end{tabular}}
\end{table}

\subsection{Force equilibrium equation}

In the complex-octonion space $\mathbb{O}$ , when the octonion force equals to zero, one can infer eight independent equilibrium equations, including the force equilibrium equation, precession equilibrium equation, mass continuity equation, and current continuity equation. Further, from the force equilibrium equation, $\textbf{N}_1^i = 0$, it is found that the force consists of the inertial force, gravity, electromagnetic force, and energy gradient and so forth. The force equilibrium equation will be influenced by the gravitational strength, electromagnetic strength, torque, and spatial dimension and so on.

According to the multiplication of octonions, the force, $\textbf{N}_1^i$ , situates in the complex-quaternion space $\mathbb{H}_g$ . Expanding the above, Eq.(3), states that the definition of force, $\textbf{N}_1^i$ , can be written as \cite{weng3},
\begin{eqnarray}
\textbf{N}_1^i = && ( W_{10}^i \textbf{g} / v_0 + \textbf{g} \times \textbf{W}_1^i / v_0
- W_{10} \textbf{b} - \textbf{b} \times \textbf{W}_1 ) / v_0
\nonumber \\
&&
+ k_{eg}^2 ( \textbf{E} \circ \textbf{W}_{20}^i / v_0 + \textbf{E} \times \textbf{W}_2^i / v_0
- \textbf{B} \circ \textbf{W}_{20} - \textbf{B} \times \textbf{W}_2 ) / v_0
\nonumber \\
&&
- ( \partial_0 \textbf{W}_1 + \nabla W_{10}^i + \nabla \times \textbf{W}_1^i ) ~,
\end{eqnarray}
where $\textbf{B}$ is the magnetic flux density, and $\textbf{E}$ is the electric field intensity. $\textbf{g}$ is the gravitational acceleration, and is relevant to the acceleration. Meanwhile $\textbf{b}$ is one part of gravitational strength, which is associated to the precessional angular velocity. And $\textbf{b}$ is called as the gravitational precessional-angular-velocity temporarily. $m_g = W_{10}^i / (k_p v_0^2)$ is the gravitational mass. $\nabla ( W_{10}^i / k_p )$ is the energy gradient. $k_p = (k - 1)$ is one coefficient, with $k$ being the dimension of the vector $\textbf{r}$ .

Especially, when the field strength $\mathbb{F}$ is weak enough, the above is approximatively degenerated into,
\begin{eqnarray}
\textbf{N}_1^i / k_p \approx  && - \partial_0 (\textbf{p} v_0)  + p_0 \textbf{g} / v_0
+ L_{10} ( \textbf{g} \times \textbf{b} + k_{eg}^2 \textbf{E} \times \textbf{B} ) / ( v_0^2 k_p )
\nonumber \\
&&
- \textbf{b} \times \textbf{p} - \nabla (p_0 v_0)
+ k_{eg}^2 ( \textbf{E} \circ \textbf{P}_0 / v_0 - \textbf{B} \times \textbf{P} ) ~,
\end{eqnarray}
where $\textbf{W}_{20}^i \approx k_p \textbf{P}_0 v_0$ . $\textbf{W}_2^i \approx \textbf{v} \times \textbf{P} $ . $( p_0 \textbf{g} / v_0 )$ is the gravity, $ \partial_0 ( - \textbf{p} v_0)$ is the inertial force, $ k_{eg}^2 ( \textbf{E} \circ \textbf{P}_0 / v_0 - \textbf{B} \times \textbf{P} ) $ is the electromagnetic force. The energy gradient is approximate to $\nabla (- p_0 v_0)$ . $p_0 \approx m_g v_0$ , $\textbf{p} = m \textbf{v}$ . $m$ is the inertial mass. $ L_{10} (k_{eg}^2 \textbf{E} \times \textbf{B} ) / ( v_0^2 k_p ) $ is in direct proportion to the electromagnetic momentum, and is relevant to the spatial dimension.

In Eq.(4), the gravitational mass, $m_g$ , consists of three parts: a) the inertial mass, $m$ ; b) the variable term, $m'$, which is relevant to the norm of field strength; c) the variable term, $m''$, which is related to the electromagnetic or/and gravitational potential energy. That is, $m_g = m + m' + m''$ . Moreover, according to the definition of term $W_{10}^i$ , the gravitational mass, $m_g$ , associates with the field strength and spatial dimension as well. It means that the potential energy and field strength both may result in the gravitational mass, $m_g$ , to deviate from the inertial mass, $m$ , to a certain extent. Certainly, there is one extreme circumstance. When the field strength and potential energy both are approximate to zero, the variable terms, $m'$ and $m''$ , will be neglected, the gravitational mass, $m_g$ , may be very close to the inertial mass, $m$ , at the time.

In the octonion field theory, the gravitational mass is variable, rather than the intrinsic property any more. This inference is close to that in the unified electro-weak theory. For the academic thought of `the mass is not an intrinsic property', one can conclude that the octonion field theory is compatible with the unified electro-weak theory, and even both of them give mutual support to each other to a certain extent. Indeed, the unified electro-weak theory goes further more. It claims even that each mass (such as, gravitational mass, inertial mass, or others) is the dynamic property, rather than the intrinsic property any more.

In the paper, the algebra of octonions should not be considered as merely one kind of useful description tool for the field theories, and the relevant research should not stop there either. Next, we cannot just satisfy to demarcate one scope of application for GR theoretically. That is, the GR is only fit for an extreme circumstance, in which the gravitational mass is invariable (see Ref.[34]). On the basis of preceding analysis, it should be utilized the dynamic property of energy gradient in the ultra-strong magnetic fields, validating the academic thought of `variable gravitational mass' in the experiments.

\begin{table}[h]
\tbl{Contrastive analysis of two energy gradients between the classical field theories and octonion field theory.}
%\center
{\begin{tabular}{@{}lll@{}}
\hline\hline
field~theory              &   energy~gradient                                     &    ultra-strong magnetic field                            \\
\hline
Newtonian~theory          &   $\textbf{N}_{B(N)} = - \nabla W_{B(N)} $            &    $\textbf{N}_{B(N)} = \textbf{F}_P + \textbf{F}_B$      \\
octonion~field~theory     &   $\textbf{N}_B = - \nabla ( p_0 v_0 + W_{B(N)} )$    &    $\textbf{N}_B = \textbf{F}_P - \textbf{F}_B$           \\
\hline\hline
\end{tabular}}
\end{table}

\section{Contrastive study}

Since a long time ago, some scholars have committed themselves to pursuing and emphasizing an objective: attempting to survey the E\"{o}tv\"{o}s experiment under more and more extreme circumstances, including the zero gravitational fields, quantum fields, free falling experiment, astronomical observation, and ultra-strong magnetic fields and so forth.

From the perspective of the ultra-strong magnetic fields, there are two types of experimental schemes to measure the variation of gravitational mass. The first experimental scheme belongs to the direct method, which measures direct a tiny variation of gravitational mass. The research thought of this scheme is simple comparatively, but it will be too difficult to implement under the existing conditions. The second experimental scheme is classified as the indirect method, which measures the energy gradient caused by the variation of gravitational mass, estimating indirect the variation of gravitational mass.

The second experimental scheme is a contrastive study, which explores the contrastive analysis between the energy gradients, $\textbf{N}_{B(N)}$ and $\textbf{N}_B$ , predicted by the Newtonian theory and octonion field theory respectively. The contrastive analysis of two energy gradients may include the magnitudes or/and directions. By all appearances, the experimental difficulty of the second experimental scheme is much less than that of the first experimental scheme. This research is beneficial to further understand the physical property of gravitational mass. And that the study has great practical significance, including the driving precisely of small mass, the explanation of movement phenomena of astrophysical jets, and the propulsion principle of some new vehicles and so forth.

With the continuous updating of the particle decelerators, the more and more particle decelerators can be applied reversely to build the new particle accelerators. The accumulation of particle decelerators drives the experimental conditions to the maturity stage, in terms of the validation of the E\"{o}tv\"{o}s experiment under the ultra-strong magnetic fields. Some relevant experiments are waiting to be let out. In the laboratories \cite{raizen}, the peak values of the magnetic strength generated by the solenoids have achieved 7.7 Tesla, which reaches to the standard (5 Tesla) of ultra-strong magnetic fields \cite{takeyama} . As a result, the energy gradients will make a great impact on the E\"{o}tv\"{o}s experiment. It is necessary to consider the influence of energy gradients on the E\"{o}tv\"{o}s experiments in the future.

\subsection{Energy gradient}

By means of the reverse application of the atomic coil-gun decelerator (see Ref.[35]), it is able to achieve one kind of particle accelerator with the ultra-strong magnetic fields. According to the classical field theories, in the external magnetic field $\textbf{B}$ generated by the solenoids, the interaction energy between the particle with the magnetic field is,
\begin{eqnarray}
W_B' = g \mu_B m_J B  ~ ,
\end{eqnarray}
where $g$ is the Lande coefficient. $\mu_B$ is the Bohr magneton. $m_J$ is the projection of the total angular momentum on the quantization axis. $B$ is the magnitude of external magnetic field $\textbf{B}$ .

Meanwhile, the energy of external magnetic field $\textbf{B}$ is,
\begin{eqnarray}
W_B'' = \int \{ B^2 / (2 \mu_e) \} dV ~ ,
\end{eqnarray}
where $dV$ is the differential element of volume.

In the atomic coil-gun decelerator, when one particle (such as, oxygen molecule, or neon atom) comes into the magnetic field generated by the solenoid from far away, the interaction energy between the particle and the magnetic field will be continuously consumed, reaching to the minimum value at the midpoint, $P_a$ , of the solenoid. Through this process, the resistance force that the particle is subjected to is,
\begin{eqnarray}
\textbf{N}_{B(N)} = - \nabla (W_B' + W_B'' ) ~.
\end{eqnarray}

As long as the particle arrives at the midpoint, $P_a$ , of the solenoid, the external magnetic field $\textbf{B}$ will be evacuated abruptly. Passing through the midpoint, $P_a$ , of the solenoid, the particle continues to move forward, but its speed is decreased, and a part of energy is lost accordingly. Such a decelerating process are repeated in subsequent solenoids again and again. After going through a series of solenoids, the speed of particle can be reduced by two orders of magnitude from the supersonic speed, approximating to zero in the end (see Ref.[35]).

Contrarily, the reverse application of the atomic coil-gun decelerator is able to accelerate the particle and achieve the thrust, constituting the propulsion principle of all-particle accelerators. In the atomic coil-gun accelerator, one particle comes into the solenoid from far away. As long as the particle arrives at the midpoint, $P_a$ , of the solenoid, the external magnetic field $\textbf{B}$ will be imposed suddenly. In case the particle reaches to the midpoint, $P_b$ , of the interval between two adjacent solenoids, the external magnetic field $\textbf{B}$ is evacuated abruptly. Passing through the midpoint, $P_b$ , of the interval, the particle achieves a part of energy, and its speed increases accordingly. Such an accelerating process can be repeated in subsequent solenoids over and over. After going through a series of solenoids, the speed of particle is capable of increasing drastically.

\subsection{Field energy}

In the existing experiments of the atomic coil-gun decelerator, the peak values of the magnetic strength generated by the solenoids reached to 7.7 Tesla. It means that the energy gradient, yielded by the energy of the ultra-strong magnetic fields, should not be neglected any more. It is necessary to consider properly the influence of this energy gradient on the particle's motion, in other stronger magnetic fields.

According to the classical field theories, in the external magnetic field $\textbf{B}$ of the solenoids, the energy of the particle and magnetic field is,
\begin{eqnarray}
W_{B(N)} = \textbf{M} \cdot \textbf{B} + \int \{ B^2 / (2 \mu_e) \} dV  ~,
\end{eqnarray}
where $\textbf{M}$ is the magnetic moment of the particle. The integrating range is mainly the space occupied by a solenoid.

From Eq.(7), in the external magnetic field $\textbf{B}$ , the energy gradient that the particle is subjected to is,
\begin{eqnarray}
\textbf{N}_{B(N)} = \textbf{F}_P + \textbf{F}_B ~,
\end{eqnarray}
where $\textbf{F}_P = - \nabla ( \textbf{M} \cdot \textbf{B} )$ , and $\textbf{F}_B = - \nabla ( \int \{ B^2 / (2 \mu_e) \} dV )$ .

From Eq.(6), the interaction energy, $\textbf{M} \cdot \textbf{B}$ , between the particle with the magnetic field is relevant to the Lande $g$ factor. In terms of different particles, the Lande $g$ factors are positive for some particles, and negative for others. Therefore, the two force terms in Eq.(10) may be added for some particles, and subtracted for others. The acceleration or deceleration cases may be partially different from each other in terms of different particles, in the ultra-strong magnetic fields.

The preceding research merely discusses the energy gradient in the classical field theories described with the vector terminology. And the following context will explore the energy gradient in the octonion field theory.

\subsection{Contrastive analysis}

According to the octonion field theory, in the external magnetic field $\textbf{B}$ generated by the solenoids, the energy density of the particle and magnetic field is $W_{10}^i$ in Eq.(4). And its energy can be written clearly as,
\begin{eqnarray}
W_B = \textbf{M} \cdot \textbf{B} + \int \{ B^2 / (2 \mu_e) \} dV - \int ( B^2 / \mu_e ) dV  ~,
\end{eqnarray}
where the last term is caused by the variable gravitational mass in the octonion field theory (see Ref.[34]).

Also the above is simplified to,
\begin{eqnarray}
W_B = \textbf{M} \cdot \textbf{B} - \int \{ B^2 / (2 \mu_e) \} dV  ~.
\end{eqnarray}

By contrast, it is found that the last term in the above has the opposite sign to that in Eq.(9). As a result, in the external magnetic field $\textbf{B}$ , the energy gradient that the particle is subjected to is (Table 2),
\begin{eqnarray}
\textbf{N}_B = \textbf{F}_P - \textbf{F}_B ~,
\end{eqnarray}
where the last term in the above has the opposite sign to that in Eq.(10).

This is a distinctive discrepancy, that is,
\begin{eqnarray}
\textbf{N}_B - \textbf{N}_{B(N)} = \nabla \int ( B^2 / \mu_e ) dV ~.
\end{eqnarray}

When the magnetic field is weak enough, the term $\textbf{F}_B$ will be much less than the term $\textbf{F}_P$ , and can be neglect in general. However, in case the magnetic field is strong enough, the term $\textbf{F}_B$ is able to exert a significant impact on the energy gradient, making a contribution to the acceleration or deceleration of the particle motions. As long as the influence of the term $\textbf{F}_B$ is strong enough to be measurable, one can check which one is correct, in terms of Eqs.(10) and (13).

It should be noted that the energy gradient, including Eq.(10), is unable to explain the movement phenomena of astrophysical jets, in the Newtonian theory. However, one equation (see Ref.[33]) similar to Eq.(13) can be considered as the dynamic of astrophysical jets, in the gravitational fields. Some astrophysical jets are capable of ejecting continuously for decades, extending to several millions of light-years in length. As one of latent applications, the propulsion principle of astrophysical jets, which is derived from the energy gradients, can be applied to design the new-style engine, that is, the astrophysical-jet engine. The later is fit for thrusting various vehicles, especially the future spaceflight vehicles. And it is similar to imitating the flight attitudes of birds, which are associated with the pressure differences, to design the existing aircrafts (Table 3).

The atomic coil-gun accelerator, which is modified from the atomic coil-gun decelerator, may be an appropriate experimental scheme, attempting to actualize the contrastive investigation of two force terms, $\textbf{N}_{B(N)}$ and $\textbf{N}_B$ , in the ultra-strong magnetic fields.

\begin{table}[h]
\tbl{As one of possible applications, the propulsion principle of astrophysical jets can be utilized to devise the new-style engine (or the astrophysical-jet engine) of the spaceflight vehicles and so forth. Similarly the flight attitudes of birds have been simulated to design the aircrafts.}
%\center
{\begin{tabular}{@{}lll@{}}
\hline\hline
imitating~object       &   working~principle                                 &    domain                       \\
\hline
birds                  &   pressure differences, caused by the               &    lift of aircrafts            \\
                       &   fluid motions                                     &                                 \\
astrophysical~jets     &   energy gradients, caused by the non-              & thrust of the engines, for      \\
                       &   uniform distribution of field strength            & spaceflight vehicles            \\
\hline\hline
\end{tabular}}
\end{table}

\section{Experiment proposal}

The energy gradient in the octonion field theory is distinct from that in Newtonian theory (see Ref.[34]). According to Eqs.(4) and (5), the energy gradient is, $\textbf{N}_B = - \nabla ( p_0 v_0 + W_{B(N)} )$,  in the octonion field theory. Meanwhile the energy gradient is, $\textbf{N}_{B(N)} = - \nabla W_{B(N)} $, in the Newtonian theory. In contrast, it is found that the energy gradients predicted by the two theories are quite different in either the magnitude or the direction. The application of appropriate experiments will be able to discriminate the discrepancy of two energy gradients predicted by the two theories.

In the laboratories, it is capable of validating directly the influence of the ultra-strong magnetic fields on the energy gradient $\textbf{N}_B$ and E\"{o}tv\"{o}s experiments. According to the predictions of the octonion field theory, a) the fluctuation of magnetic strength will vary the gravitational mass, when the ultra-strong magnetic field is the uniform distribution; b) the fluctuation of magnetic strength will not only change the gravitational mass, but also induce the energy gradient, when the ultra-strong magnetic field is the non-uniform distribution, for example, the pulsed magnetic fields. As a result, not matter whether the distribution is uniform or not, the emergence of the ultra-strong magnetic fields must destruct the preceding equilibrium state of force, shifting the neutral test particle to another equilibrium point in the E\"{o}tv\"{o}s experiments. However, each of the existing E\"{o}tv\"{o}s experiments has never been inspected under any circumstance of the ultra-strong magnetic fields until now. Apparently, it is necessary to verify E\"{o}tv\"{o}s experiment in the ultra-strong magnetic fields as soon as possible.

The existing sophisticated laboratory conditions mean the difficulty of measuring the tiny variation of gravitational mass directly. Presenting a striking contrast to the above is that the determination of energy gradient is capable of reducing the difficulty of the experiment to a great extent. That is, one can explore the contrastive study of the magnitudes (and directions) between the two energy gradients, $\textbf{N}_B$ and $\textbf{N}_{B(N)}$ . This is an effective indirect method. By all appearances, it is able to improve and modify the experimental equipments of the atomic coil-gun decelerator, meeting the requirement of the atomic coil-gun accelerator, especially optimizing the distribution of ultra-strong magnetic fields, and modulating the flight time of particles within the magnetic fields (see Ref.[35]).

\subsection{Reverse application}

At present, there have been a variety of particles decelerators, reducing observably the velocity of some molecules or atoms, including the charged or neutral particles. For instance, considering the energy gradient as the dynamic, it is capable of building the working principle of one particle decelerator. Reversely applying some working principles of these existing particle decelerators, one can achieve a few new particle accelerators, actualizing the validation of E\"{o}tv\"{o}s experiment under the ultra-strong magnetic fields ultimately.

There are two types of particle decelerators to exploit the electromagnetic drive. One type of particle decelerator is driven by the magnetic fields, while the other by the electric fields. Both of them utilize the energy gradient as the thrust. The particle decelerators, driven by the magnetic fields, make use of the interaction of polar particles with the magnetic fields, that is, Zeeman Effect. The particle decelerators, driven by the electric fields, employ the interaction of polar particles with the electric fields, that is, Stark Effect.

The paper discusses mainly the particle decelerators driven by the magnetic fields, and its reverse application to the particle accelerator, that is, the atomic coil-gun accelerator. The particle decelerators, driven by the magnetic fields, utilize the energy gradient as the dynamic, decelerating the particles (atoms or molecules). In the magnetic fields, the energy mostly consists of the interaction energy between the magnetic field with the magnetic moment of the particle, and the energy of magnetic field itself. Further, the particle decelerators, driven by the magnetic fields, are divided into two parts, Zeeman decelerator and atomic coil-gun decelerator. And we just explore the atomic coil-gun decelerator driven by the magnetic fields. This type of decelerator is capable of decelerating some particles, including the paramagnetic atom/molecule, ferromagnetic atom/molecule, alkali atom/molecule, lithium hydride molecule, metastable nitrogen atom, metastable helium atom, neutral atom/molecule and so forth. Certainly, there may be a few overlaps among the above sorts of particles. Because a majority of elements are paramagnetic, the purpose of the atomic coil-gun decelerator is very extensive, including the deceleration of the neon, oxygen, hydrogen, and deuterium and so on.

If the atomic coil-gun decelerator is utilized reversely, it is able to accelerate the most of the paramagnetic particles under the ultra-strong magnetic fields, especially the nitrogen atom, oxygen molecule, hydrogen molecule, and neon atom and others.

\subsection{Proposed scheme}

By means of the atomic coil-gun accelerator, the energy gradient is cable to accelerate the particles under the ultra-strong magnetic fields. a) In the atomic coil-gun accelerator, one particle comes into the solenoid from far away. As long as the particle arrives at the midpoint, $P_a$ , of the solenoid, the external magnetic field $\textbf{B}$ will be imposed suddenly. The interaction energy between the particle and the magnetic field will be continuously accumulated, reaching to the maximum value at the midpoint, $P_b$ , of the interval between two adjacent solenoids. In case the particle arrives at the midpoint, $P_b$ , the external magnetic field $\textbf{B}$ is evacuated abruptly. Passing through the midpoint, $P_b$ , of the interval, the particle achieves a part of energy, and its speed increases accordingly. b) The distribution of the ultra-strong magnetic fields is gradient and symmetric, utilizing the battery series to be the power supply. After going through a series of solenoids, the speed of particle is capable of increasing drastically. c) Making a comparison of particles' speeds between the speed selection and micro-channel plate (or quadrupole mass spectrometer and others), it is capable of analyzing the acceleration effect of the accelerator on the particles.

The part of this study mainly consists of three components: the polarized beams of particles, the arrangement of flight times, and the ultra-strong magnetic fields with the gradient distribution. The experimental equipments include the polarized particle source, speed selection, skimmer, accelerated coils, data processing system, and electrical control system and so forth. Obviously, the research is very important on the ultra-strong magnetic fields with the gradient distribution.

The polarized beam of particles goes through the adjustment electric field and the speed selection with a dual-slit collimated, achieving the required incident speed. Next, the beam of particles passes through the skimmer, allowing a few selected particles to enter the accelerator. After experiencing the continuously acceleration in a series of ultra-strong magnetic fields with the gradient distribution, the emergent particles comes into the micro-channel plate (or quadrupole mass spectrometer and others), measuring the speed of the emergent particles.

In a word, making use of the researches relevant to the atomic coil-gun accelerator (see Ref.[35]), it is able to construct and evolve the proposed scheme and experimental facilities, for the `contrastive analysis of two energy gradients' in the ultra-strong magnetic fields. In the proposed scheme, the paper consults either the design proposal or the reverse application about the mass spectrograph, polarized particle source, Zeeman decelerator, and Stark decelerator and so forth, improving the experiment proposal in a certain extent.

\section{Conclusions and discussions}

It is well known that the mass has been transformed from the intrinsic property into the dynamic property, according to the unified electro-weak theory. A complete and profound understanding of this viewpoint will undoubtedly challenge the results and inferences of the E\"{o}tv\"{o}s experiments, revealing a few new physical properties of the mass. Because the measurement method of the gravitational mass is disparate from that of the inertial mass, the scholars have been suspecting the equivalence between the two masses for a long time. Up to now each one of E\"{o}tv\"{o}s experiments is merely validated experimentally in the quite weak gravitational fields, but it has never been verified in the comparatively strong magnetic fields or gravitational fields. This limitation is aggravating the existing serious qualms about the equivalence of two masses. As a result, it is necessary to survey experimentally the equivalence of two masses in the ultra-strong magnetic fields.

Making use of the complex octonions rather than split-octonions \cite{negi1, chanyal1}, it is able to describe simultaneously the physical quantities of the gravitational and electromagnetic fields, including the field potential, field strength, field source, linear momentum, angular momentum, torque, and force and so forth. When the octonion force is equal to zero, one can achieve eight independent equilibrium equations, especially the force equilibrium equation, precession equilibrium equation, mass continuity equation, and current continuity equation. From the force equilibrium equation, it is capable of defining the inertial mass and gravitational mass respectively. In the octonion field theory, the gravitational mass is variable. The gravitational mass is the sum of the inertial mass and a few tiny terms. And that the tiny terms are varied with the fluctuation of either the field strength or potential energy. In the experiments, it is not an easy thing to measure directly the tiny variations of the tiny terms. In contrast to this, it must be comparatively easy to measure the energy gradient concerned with the tiny variations of these tiny terms.

The prediction of the energy gradient in the octonion field theory is in sharp contrast to that in the Newtonian theory. The energy gradient predicted by the Newtonian theory is $\textbf{N}_{B(N)}$ , while that by the octonion field theory is $\textbf{N}_B$ . If the magnetic field is weak enough, the discrepancy of two force terms, $\textbf{N}_{B(N)}$ and $\textbf{N}_B$, can be neglected. However, under the ultra-strong magnetic fields, the differentia of two force terms, $\textbf{N}_{B(N)}$ and $\textbf{N}_B$ , may be acute, including the magnitude and direction. It is necessary to develop the contrastive study of two force terms, $\textbf{N}_{B(N)}$ and $\textbf{N}_B$ , helping us to check which theory is correct. One of experiment schemes is to consult and improve the reverse application of the atomic coil-gun decelerator, accelerating the paramagnetic particles and so on in the ultra-strong magnetic fields, measuring the accelerating effect for the particles in the magnetic fields with the gradient distributions.

It should be noted that the paper explored only some simple cases for the contributions of the ultra-strong magnetic fields on the gravitational mass and energy gradient. Despite its preliminary characteristics, this study can clearly indicate that the external ultra-strong magnetic fields can result in the gravitational mass of one particle to deviate from the inertial mass to a certain extent. And that the external ultra-strong magnetic fields will cause the energy gradient, $\textbf{N}_B$ , predicted by the octonion field theory to depart from the energy gradient, $\textbf{N}_{B(N)}$ , predicted by the Newtonian theory. This affords us an opportunity to check the E\"{o}tv\"{o}s experiment in the ultra-strong magnetic fields. In the following study, it is going to further research other possible influence factors for the gravitational mass from the force equilibrium equation theoretically. In the experiments, we will apply the energy gradient to measure the impact of the magnetic flux density $\textbf{B}$ on the gravitational mass, seeking further a few new physical properties of gravitational mass on the basis of the variable gravitational mass.

\section*{Acknowledgments}
The author is indebted to the anonymous referees for their valuable comments on the previous manuscripts. This project was supported partially by the National Natural Science Foundation of China under grant number 60677039.

%\begin{thebibliography}{000} %for 3 digits
%\begin{thebibliography}{00}  %for 2 digits

\end{document}